\documentclass[12pt]{article}
\begin{document}
\title{Generalized Quantum Theory, Contextual Emergence and Non-Hierarchic Alternatives}
\author{Hartmann R\"omer \\Institute of Physics, University of Freiburg, Germany
\\http://omnibus.uni-freiburg.de}
\date{March 23rd 2015}
\maketitle \pagenumbering{arabic}

\begin{abstract}
The concept of emergence is critically analyzed in particular with
respect to the assumed emergence of mental properties from a
neuronal basis. We argue that so-called contextual emergence is
needed to avoid an eliminatory reductionism. Quantum-like features
of the emergent qualities are to be expected. As a consequence,
non-causal relations like entanglement correlations have to be
considered as full fledged elements of reality. "Observable
extension" is proposed as a contextual alternative to emergence
avoiding the asymmetry between purportedly basic and emergent
properties.
\end{abstract}

%\tableofcontents
\section{Introduction}
{\tt By convention sweet, by convention bitter, by convention hot,
by convention cold, by convention color: but in reality atoms and
void.} This is the first and for all times prototypal formulation
of a reductionist world view given in the fifth century B.C. by
Democritus from Abdera. The endeavor is understanding the world in
terms of a limited number of ''primary qualities'' of a basic
layer of reality like positions and velocities of atoms and
reducing ''secondary qualities'' of a somehow subordinate
ontological status like color and smell to the primary ones. For
good reasons, some version of physical reductionism is a
widespread if not dominant attitude in contemporary science. It
offers an attractive reduction of complexity in understanding
large parts of our world and it takes profit from the impressive
success of physics in  exactness, certainty, coherence and
applicability. The reduction of thermodynamics to mechanics is
often considered as paradigmatic for the success of a reductionist
program. (But see our discussion in Section 3.) \emph{Neuronal
reductionism} as a strategy of understanding mental phenomena in
terms of neuronal activities has many vigorous proponents. This is
another example of a physical reductionism, because the
possibility to understand
neuronal activity in physical terms is generally accepted.\\
\emph{Eliminative reductionism} is an extreme form of reductionism
attributing reality only to the basic layer. For instance,
eliminative neural reductionism \cite{churchland} attributes to
the ''popular psychology'' terminology only the meaning of an
incomplete shorthand notation for the true and exact neuronal
description. This radical view is rarely adopted and will not be
discussed further. Much more widespread is \emph{emergentism}, an
attitude granting to notions of the secondary, ''higher'',
''emergent layer'' its own although ontologically somehow
subordinate status. The claim is that systems described in terms
of the basic layer will develop new and surprising features once a
certain threshold of complexity is passed. Many versions of
emergentism are advocated reaching from a milder form of
reductionism up to a decidedly anti-reductionist attitude. In more
formal terms, the question of the interpretation of emergentism
becomes a problem of the specification of the relationship between
two different descriptions or modelisations of a part of reality,
one of them considered ''basic''and one considered ''emergent''.
Three questions are of particular interest in this context. (1)
What is the ontological status of the emergent layer? (2) What is
the novelty status of the emergent layer? (3) What about the
possibility of ''downward causation'' from the emergent to the
basic layer. Approaching these questions we shall proceed as
follows: (More material can be found in \cite{Roemer2013}.)\\
First we shall present a very general formal scheme for describing
and modelling systems of most general type. It has been developed
under the name of ''Weak'' or \emph{Generalized Quantum
Theory}(GQT)  \cite{ARW, AFR2006, FR2010} arising from physical
quantum theory by shedding off all formal features pertaining to
physics in the narrow sense and thus widening the range of
applicability beyond physics still keeping salient quantum notions
like complementarity and entanglement. GQT can be seen to be in
direct accordance with basic categorial fundamentals of the human
cognitive system. GQT strongly suggests that quantum-like features
of systems are generic and that the ontological constitution of
systems in classical physics should be considered exceptional.
This message is noteworthy, because neuronal emergentism is
normally inspired by a classical or even mechanistic world model.\\
Next we shall critically analyze the standard example of the
emergence of thermodynamics and arrive at the notion of
\emph{Contextual Emergence} \cite{BishopAtmanspacher2006,
atmanspacherbeimgraben}. Then we shall investigate to what extent
neuronal emergence fulfills the
criteria of Contextual Emergence.\\
Finally, we propose \emph{observable extension} as a more
symmetric alternative to the hierarchical concept of emergence and
give tentative answers to the three questions raised above.

\section{Generalized Quantum Theory}
World is never given to us directly but primarily only as it
appears on our internal stage. Naive realism assumes that the
world, at least in essence, really is like it appears to us. This
is a very strong  assumption underrating the active role of the
human cognitive system and the human activity as a ''model
builder''. The preferred world model of contemporary physical and
neuronal reductionism exhibits the influence of classical
mechanics and is
in danger to confuse a world model with the world itself.\\
In fact the appearance of the world is bound  to fundamental
\emph{existentials}, constitutive features of our human mode of
existence and thereby also categorial elements
 of human cognition,
similar to Kant's categories. Four of them are of particular
importance for us: (For more details see \cite{Roemer2011b})
\begin{enumerate}
\item \emph{Excentricity/oppositeness}: Every perception or
cognition of the form accessible to us is irrevocably bound to the
figure of oppositeness by always being the perception or cognition
of something by someone. The \emph{epistemic cut} separating the
''observer'' from the ''observed object'' may be movable but it is
never completely removable.
 \item \emph{Temporality}: The world is not given to us in the mode of a timeless panoramic
 picture but rather like a movie or a temporal sequence of events which occur in the running
 window of a distinguished ''now''.
 \item \emph{Facticity}: We do not so much live in a world of
 potentialities but rather in a world of facts, which hit on us
 and occur to us. The ''now'' is of prototypal facticity.
\item  \emph{Freedom and causality} are not in contradiction but
offshoots from the same root of temporality differentiated into
past, present and future. They rely on each other: Causality is
required for freely chosen actions to have predictable effects and
causal regularities are detected by freely creating causes and
observing their effects.
\end{enumerate}

Weak or \emph{Generalized  Quantum Theory} (GQT) \cite{ARW,
AFR2006, FR2010} is a conceptual core of quantum theory, which
arose from an axiomatic formulation of physical quantum theory by
leaving out all features which seemed to be special to physical
systems. As we shall see, it takes into account the above
existential universals right in the heart of its structure. GQT is
not physics but a very general theory of the structure of observed
systems. For the purposes of this note, it suffices to give a
short account of the vital structural features of GQT.
For recent developments and applications see \cite{FR2010, AR2012}. \\
\vspace{0.2cm}
The following notions are taken over from quantum physics:\\
\emph{System}: A system is anything which can be (imagined to be)
isolated from the rest of the world and be subject to an
investigation. A system can be as general as ''impressionism'', a
school of art together with all persons involved in production and
interpretation. Unlike the situation in, e.g., classical mechanics
the identification of a system is not always a trivial procedure
but sometimes a creative act. In many cases it is possible to
define \emph{subsystems} inside a system.
\\[0.2cm]
\emph{State}: A system must have the capacity to reside in
different states without losing its identity as a system. One may
differentiate between \emph{pure states}, which correspond to
maximal possible knowledge of the system and \emph{mixed states}
corresponding to incomplete knowledge. \\[0.2cm]
\emph{Observable}: An observable corresponds to a feature of a
system, which can be investigated in a more or less meaningful
way. \emph{Global observables} pertain to the system as a whole,
\emph{local observables} pertain to subsystems. In the
above-mentioned example, observables may, for instance, correspond
to esthetic investigations.
\\[0.2cm]
\emph{Measurement}: Doing a measurement of an observable $A$ means
performing the investigation which belongs to the observable $A$
and arriving at a result $a$, which can claim \emph{factual
validity}. What factual validity means depends on the system:
Validity of a measurement result for a system of physics, internal
conviction for self observation, consensus for groups of human
beings. The result of the measurement of $A$ will in general
depend on the state $z$ of the system before the measurement but
will not be completely determined by it. \vspace{0.4cm}

Immediately after a measurement of an observable $A$ with result
$a$ , the system will be in an \emph{eigenstate} $z_a$ of the
observable $A$ with \emph{eigenvalue} $a$. The eigenstate $z_a$ is
a state, for which an immediate repetition of the measurement of
the same observable $A$ will again yield the same result $a$ with
certainty, and after this repeated measurement the system will
still be in the same state $z_a$. This property, which is also
crucial in quantum physics justifies the terminology ``eigenstate
of an observable $A$'' for $z_a$ and ``eigenvalue'' for the result
$a$. We emphasize that this is an idealized description of a
measurement process abstracting from its detailed temporal structure.\\
Two observables $A$ and $B$ are called \emph{complementary}, if
the corresponding measurements are not interchangeable. This means
that the state of the system depends on the order in which the
measurement results, say $a$ and $b$, were obtained. If the last
measurement was a measurement of $A$, the system will end up in an
eigenstate $z_a$ of $A$, and if the last measurement was a
measurement of $B$, an eigenstate $z_b$ will result eventually.
For complementary observables $A$ and $B$ there will be at least
some eigenvalue, say $a$, of one of the observables for which no
common eigenstate $z_{ab}$ of both observables exists. This means
that it is not generally possible to ascribe sharp values to the
complementary observables $A$ and $B$, although both of them may
be equally important for the description of the system. This is
the essence of quantum theoretical complementarity which is well
defined also for GQT. Some people prefer to apply the term
"complementarity" to what we call "maximal complementarity":
Knowledge of the value od $A$ implies complete indeterminacy of
the value of $B$ and voce versa. This holds for the position and
momentum observables $Q$ and $P$ of quantum mechanics. But
according to this definition
already oblique components of angular momentum would not be complementary.\\
Notice, that a measurement will in general change the state of a
system by production of an eigenstate of the last-measured
observable. This generic quantum feature is realized in a
paradigmatic way for the human mind under the first person
perspective of self-observation. It will also hold for human
thought in general  and for all kinds of discourse, belief or
decision systems. Quantum features like complementarity and
indeterminacy should be common there. Detailed empirical
investigations of quantum features in psychological systems have
been performed for bistable perception \cite{AFR2004, ABFKR,
AFR2008}, human thought and the formation of concepts
\cite{AertsGaboraSozzo, AertsSozzoVeloz}, apparent irrationality
and non-classical logic in decision processes \cite{AertsDHooghe,
PothosBusemeyer}, semantic networks, learning and order effects in
questionnaires \cite{WangBusemeyer, AR2012}. For further
information see \cite{QuantumConsciousness} There are striking
similarities of the measurement process with creative processes
\cite{RoemerJacoby2011}.\\
 Non-complementary observables, for which the order of
measurement does not matter, are called \emph{compatible}. After
the measurement of compatible observables $A$ and $B$ with results
$a$ and $b$, the system will be in the same common eigenstate
$z_{ab}$ of $A$ and $B$ irrespective of the order in which the
measurements were performed. In classical systems all observables
are compatible and possess simultaneous eigenstates, and the
phenomenon of complementarity does not occur. It should be clear
from our general structural consideration and from the examples
given, that this is a strong additional assumption. For general
reasons, quantum-like behavior
of systems should be the rule rather than the exception.\\
We also see, how the above-mentioned categorial universals are
built into the structure of GQT \cite{Roemer2011b}:\\
Excentricity is taken into account by the pivotal position of
measurement in GQT. In physical quantum theory the epistemic cut
is known under the name of Heisenberg cut. Moreover, observables,
right by definition, assume the existence of an epistemic cut.
They are sitting right astride on it with a footing both on the
side of  the observer and the observed. Temporality is present in
the relevance of the (temporal) ordering of measurements. In
addition, quantum systems in general have temporal dynamics.
Facticity resides in the factual character of measurement results.
Freedom and causality show up in GQT in the strange interplay of
freedom in the choice of the observable to be measured and the
causal dynamics of the system.

\emph{Entanglement} can also be defined in the framework of
Generalized Quantum Theory \cite{ARW, AFR2006, FR2010,
Roemer2011}. It may and will show up under the following
conditions:\\
\begin{enumerate}
\item Subsystems can be identified within the system such that
local observables pertaining to different subsystems are
compatible.

\item There is a global observable of the total system, which is
complementary to local observables of the subsystems.

\item The system is in an \emph{entangled state} for instance in
an eigenstate of the above-mentioned global observable and not an
eigenstate of the local observables.
\end{enumerate}

Given these conditions, the measured values of the local
observables will be uncertain because of the complementarity of
the global and the local observables. However, so-called
\emph{entanglement correlations} will be observed between the
measured values of the local observables pertaining to different
subsystems. These correlations are non-local and instantaneous.
They are not usable for signals or causal influences. They are
non-causal order structures resulting from the holistic structure
of quantum systems. The crucial importance  of non-causal ordering
structures is a fundamental message of quantum theory. The
explanatory monopole of causal relations, often tacitly assumed
under the influence of a mechanical paradigm
cannot be held up.\vspace{0.4cm} \\
Comparing Generalized with full physical quantum theory the
following vital differences are worth noticing:
\begin{itemize}

\item In its minimal version and in contrast to other approaches
\cite{Aertsetal}, GQT does not ascribe quantified probabilities to
the outcomes of measurements of an observable $A$ in a given state
$z$. Indeed, to give just one example, for esthetic observables
quantified probabilities seem to be inappropriate from the outset.
What rather remains are modal logical qualifications like
``impossible'', ``possible'' and ``certain''. Related to the
absence of quantified probabilities, the set of states in GQT is
in general not modelled by a linear Hilbert space.

\item Related to this, GQT in its minimal form provides no basis
for the derivation of inequalities of Bell's type for measurement
probabilities, which allow for the conclusion that the
indeterminacies of measurement values are of an intrinsic ontic
nature.  In many (but not all) applications of GQT indeterminacies
may be epistemic and due to incomplete knowledge of the full state
or uncontrollable perturbations by outside influences or by the
process of measurement. Notice that complementarity in the sense
of GQT may even occur as a consequence of different partitions in
coarse grained classical dynamical systems \cite {GA2006,BFA2013}.
In this sense, GQT is a phenomenological framework theory allowing
to leave the question of the ontic or epistemic character of
indeterminacies open.
\end{itemize}
For some applications (see, e.g., \cite{AFR2004, ABFKR, AFR2008,
WangBusemeyer, AR2012}, ) one may want to enrich the
above-described minimal scheme of GQT by adding further structure,
e.g., an underlying Hilbert space
structure for the states. \\

\section{Contextual Emergence}
The statistical theory of thermodynamics is often considered to be
the classic example of a successful reduction and a
well-understood emergence relationship. We already mentioned that
emergence concerns the relationship between two descriptions of a
part of the world, one of them primary and basic, one secondary
and emergent. In this exemplary case the basic level is a  system
of (many) particles described microscopically by classical or
quantum mechanics with the positions, momenta and spins of the
particles as fundamental observables. The emergent macroscopic
level is described as a system of classical constitution with
different macroscopic observables like volume, pressure,
temperature and entropy. The macroscopic state is determined by
the values of a sufficient number of macroscopic observables. The
macroscopic observables \emph{supervene} \cite{Kim1993} the
microscopic observables because a change in the macroscopic  ones
is necessarily accompanied by a change in the microscopic ones but
nut necessarily vice versa. Both the macroscopic and the
microscopic description are formally  well developed complete
theories making emergence and supervenience exemplarily clear
issues in this case.\\
On closer inspection the reduction of thermodynamics to
microphysics proceeds in two steps:\\
First, the detailed microscopic description of states, which is
neither feasible nor even desirable for a large system is replaced
by a statistical description. This is done by first introducing
\emph{mixed states}, i.e. ensembles or sets of pure microscopic
states with an attribution of a probability  to each of them. In a
second step, \emph{macrostates}, defined by the values of
macroscopic thermodynamical observable are identified with
appropriate mixed states. There are many mixed states without
thermodynamic interpretation.\\
Notice, that the reduction of thermodynamics to microphysics is
not to be understood such that in complex microphysical systems
after passing a complexity threshold  completely new concepts of a
thermodynamic description arise automatically by themselves and
from nothing. The concept of probability, applied in the first
step is not newly born but pre-existent. It is also applicable to
small microscopic systems for which, as opposed to large systems,
a
detailed microscopic description is still feasible. \\
Some observables like the total energy are common to the
microscopic and macroscopic description. In general, the
identification of macroscopic observables like volume and
temperature is neither enforced  by the microscopic description
nor by the concept of mixed states. It comes about by applying
different contexts to a section of reality in addition to its
microphysical description. For this reason, Atmanspacher, Bishop
and beim Graben \cite{BishopAtmanspacher2006,
atmanspacherbeimgraben} talk
about \emph{Contextual Emergence}.\\
For a successful Contextual Emergence relationship a further
condition must be fulfilled: The mixed states corresponding to
thermodynamic macrostates must have a sufficient degree of
stability under the microscopic dynamics. Otherwise, these mixed
states would quickly develop into mixed states without
thermodynamic interpretation. A change of the topology of the states
may be necessary to achieve stability.\\
Thermodynamics is a particularly clear case of Contextual
Emergence. There is a widespread hope, that other emergence
situations conform with this example. This amounts to demanding a
lot: Both the basic and the emergent layer must be well formalized
and endowed with dynamics which meet the above-mentioned stability
requirement. \\
As for the emergence of a mental from a neuronal description the
situation seems to be as follows: For the neuronal layer, a
satisfactory formal description is largely available. States and
observables are essentially under control, perhaps with some
restrictions for the dynamics of larger neuronal assemblies. The
situation is much more problematic on the mental level. First of
all, we  expect it to be organized in a quantum-like rather than a
classical manner \cite{QuantumConsciousness, RW2011}, making
reduction to neuronal properties much more difficult. Moreover,
there is no comprehensive description and classification of mental
states and observables. The dynamics on the mental level can in no
way considered to be understood. Exactness is sometimes attempted
by restriction to a small set of mental observable, which are,
however, so much devised with an eye on the neuronal substrate
that reduction becomes almost tautological. The status of the
stability requirement is also unclear. Mental dynamics is largely
unknown, and one has the impression that on the one hand  quite
different neuronal states frequently correspond to similar mental
states and on the other hand sometimes a small change of the
neuronal state often leads to a large change of the mental state.
Mental and neuronal properties will correspond to very different
topologies and partitions of states. It is just this difference,
which makes the "emergence" of a quantum-like mental  from a
classical neuronal domain possible.
\\
In any case, contextuality will be decisive for neuronal
emergentism at least as much as for thermodynamics. The hope for
completely reductionist emergentism for the mental domain deriving
all mental features in an automatic and cogent way from neuronal
ones seems to be futile.
\section{Alternatives to Emergence}
Emergence and supervenience are genuinely asymmetric concepts
distinguishing between a basic, lower, ontologically primary and
an emergent, higher, ontologically secondary level. The vision of
physical reductionism is a hierarchic structure of the world with
a basic physical layer given, for instance, by elementary particle
physics and above it a tower of higher stepwise emergent levels
like chemistry, life and mind. One may ask oneself, whether such a
hierarchic ordering of the world really has an ontological status
reflecting a real feature of the world or whether it is epistemic
and arises only from a particular description of the world such
that in a different description the layers and/or their ordering
might be different.
\\ A frequently invoked argument in favor of an ontological hierarchy
is concerned with complexity.  Emergence arises, when the
complexity on the basic layer exceeds a certain threshold.
However, one should keep in mind that complexity is an epistemic
notion referring to a mode of description. What is complex in one
description may be simple in another description. Consider the
exemplary case of the emergence of thermodynamics from microscopic
mechanics. Experience shows that for thermodynamical systems far
from equilibrium the treatment of the fluctuating behavior of the
thermodynamic observables becomes complicated to the verge of
intractability. In this situation, a microscopic atomic
description suggests itself and, indeed,  molecular dynamics is
the method of choice here. In this case, the direction of
emergence seems to be reversed. As a matter of fact, thermodynamic
fluctuations were historically one important reason for the
acceptance of atomism.\\
Contextuality of emergence and the high degree of autonomy of the
emergent level are further arguments against an ontological
hierarchy. As an example, the function of a software is largely
independent of its underlying hardware substrate realization and
could be achieved in many different ways. It is also conceivable
that mental properties could rest on a basis quite different from
neurons. As far as physics is concerned, it is quite uncertain,
whether a really fundamental level has already been reached and
whether it exists at all. In addition, recent developments seem to
indicate that elementary particle theory is not understandable
without and closely interwoven with cosmology, which according to
the traditional view should emerge from particle theory.\\
From the preceding considerations we see, that the ontological
status of the emerging layer is quite strong and independent and
not strictly subordinate. Moreover, an ontologically hierarchical
order of the world is disputable. In this situation it is
suggestive to question the asymmetry inherent in the concept of
emergence with the distinction between basic and emergent layer.
Emergence is a relationship between two formal systems with
different sets of states and observables both of them describing a
certain sector of reality. Neural emergence with a basic neuronal
and an emergent mental layer is the example of central interest
for us. A more symmetric alternative to emergence in general and
neural emergence in particular would be \emph{extension of the set
of observables} or more briefly \emph{Observable Extension}. This
amounts to describing a sector of reality by just one formal
system with one large set of observables corresponding to both
layers of emergence. For example, rather than neuronal emergence,
one considers just one large comprehensive system "man" or even
"man plus physical and social environment" which contains both
neuronal and mental observables. This is in accordance with a
standpoint of neutral dual aspect monism with respect to the
matter-mind problem. We already saw that complementarity should
occur between mental observables of the comprehensive system. One
can also argue that complementarity between neuronal observables
on one side and mental observables on the other side should be
common \cite{RW2011}. So, in any case, the comprehensive
matter-mind system should be quantum-like in the sense of GQT.\\
For ''Observable Extension'' contextuality is at least as vital as
for Contextual Emergence. The identification of additional
observables is not automatically enforced by the other observables
but corresponds to the introduction of new concepts
and contexts into the investigation of the comprehensive system.\\
For complementary observables different values for one of them
will in general change expectations for the measured values of the
other one. So, the basic dictum of supervenience "Change on the
emergent layer leads to change on the basic layer" also holds for
complementary observables in an appropriately symmetrized form.

\section{Conclusions}
We are now in the position to attempt answers to the three
questions raised in the Introduction.\\
The first two questions concerned the ontological status and the
novelty of "emergent" properties. It should be abundantly clear
from the preceding considerations, in particular from the analysis
of the paradigmatic case of the emergence of thermodynamics, that
contextuality is a vital element in the discussion of emergence.
It is even more central in the symmetric alternative concept of
"Observable Extension". Unless one is willing to adopt a radical
eliminatory reductionism, the "emergent" properties are not
automatically generated by the "basic" layer beyond a certain
threshold of complexity. Rather they come from different contexts
becoming applicable or useful. Finding such contexts and detecting
their applicability is a subtle achievement of creativity whose
origin is a deep and difficult question \cite{RoemerJacoby2011}.
As for the novelty of the "emergent" features: They are not
suddenly born from the basic layer like Athena from the head of
Zeus but they correspond to preexistent notions and the novelty
consists in their applicability or usefulness with increasing
complexity on the basic level. To give a very simple example: If
the complexity of a system of points in a plain is increased from
two to three points the concept of angles becomes applicable and
useful, but it was already existent and not newly born with the
appearance of the
third point.\\
The third question raised in the Introduction was about ''downward
causation'' from the emergent to the basic level. It is formulated
in a slightly provocative way as \emph{Kim's Dilemma}
\cite{Kim2003}. For the status of the emergent mental level in
relation to its neuronal basic the following dire alternative
seems to hold: Either mental properties are just abbreviations for
neural properties in the sense of an eliminative reductionism or
else, due to  the assumed causal closure of the physical world,
they are impotent and causally decoupled from the physical world.
This leads Kim to the assertion that emergence and supervenience
are formulations rather than solutions of a
problem.\\
The assumption of causal closure of the physical world is
questionable \cite{Roemer2013}. But even taking it for granted, a
smooth resolution of the dilemma comes from the concept of
contextuality both in the form of Contextual Emergence and of
"Observable Extension". There is no causal interaction between the
neuronal and the mental level and, in fact, no such interaction is
needed. The relationship of the different layers is not causal in
its nature but a correspondence and order structure, which is due
to the fact that the same part of reality is observed from
different perspectives. From the example of the emergence of
thermodynamics we easily see that the relationship between the
microscopic and thermodynamical description is not a causal one.
Of course, the microstate changes when the values of thermodynamic
variables change, but this simultaneity in change only reflects
the fact, that both descriptions are different sides of the same
medal. This becomes even clearer if we consider complementary
observbles in one and the same system. Nobody would interpret the
subtle relationship between position and momentum distributions as
causal effects.\\
Kim's dilemma just results from an unjustified monopolization of
causal relationships as  explanatory structures. Quantum theory
with its inevitable  non-causal entanglement correlations lends
yet another disproof of such a one-sided claim.

%\bibliography{Literaturverzeichnis}

\begin{thebibliography}{10}

\bibitem{churchland}
P.M. Churchland.
\newblock {\em The Engine of reason, the Seat of the Soul}.
\newblock Cambridge University Press, 1995.

\bibitem{Roemer2013}
H.~R{\"o}mer.
\newblock Emergenz und {E}volution.
\newblock {\em Talk {O}ffenburg, {O}ctober 2013, available on the author's
  homepage, to be published}, 2013.

\bibitem{ARW}
H.~Atmanspacher, H.~R\"omer, and H.~Walach.
\newblock Weak quantum theory: Complementarity and entanglement in physics and
  beyond.
\newblock {\em Foundations of Pysics}, 32:379--406, 2002.

\bibitem{AFR2006}
H.~Atmanspacher, T.~Filk, and H.~R\"omer.
\newblock Weak quantum theory: Formal framework and selected applications.
\newblock In G.~Adenier, A.~Yu. Khrennikov, and T.M. Nieuwenhuizen, editors,
  {\em Quantum Theory: Reconsiderations and Foundations}, pages 34--46.
  American Institute of Physics, New York, 2006.

\bibitem{FR2010}
T.~Filk and H.~R\"omer.
\newblock Generalized quantum theory: Overview and latest developments.
\newblock {\em Axiomathes}, 21,2:211--220; DOI 10.1007/s10516--010--9136--6,
  2011, http://www.springerlink.com/content/547247hn62jw7645/fulltext.pdf.

\bibitem{BishopAtmanspacher2006}
R.~Bishop and H.~Atmanspacher.
\newblock Contextual emergence in the description of properties.
\newblock {\em Foundations of Physics}, 36:1753--1777, 2006.

\bibitem{atmanspacherbeimgraben}
H.~Atmanspacher and P.~beim Graben.
\newblock Contextual emergence of mental states from neurodynamics.
\newblock {\em Chaos and Complexity Letters,}, 2:151--168, 2007 (see also
  Scholarpedia-article www.scholarpedia.org/article/contextual\_emergence ).

\bibitem{Roemer2011b}
H.~R{\"o}mer.
\newblock Why do we see a classical world? http://arxiv.org/abs/1112.6271.
\newblock {\em Travaux Math\'{e}matiques}, XX:167--186, 2012.

\bibitem{AR2012}
H.~Atmanspacher and H.~R\"omer.
\newblock Order effects in sequential measurements of non-commutative
  psychological observables, http://arxiv.org/abs/1201.4685.
\newblock {\em Journal of Mathematical Psychology}, 56:274--280, 2012.

\bibitem{AFR2004}
H.~Atmanspacher, T.~Filk, and H.~R\"omer.
\newblock Quantum {Z}eno features of bistable perception.
\newblock {\em Bilogical Cybernetics}, 90:33--40, 2004.

\bibitem{ABFKR}
H.~Atmanspacher, M.~Bach, T.~Filk, J.~Kornmeier, and H.~R\"omer.
\newblock Cognitive time scales in a {N}ecker-{Z}eno model of bistable
  perception.
\newblock {\em The Open Cybernetics and Systemic Journal}, 2:234--251, 2008.

\bibitem{AFR2008}
H.~Atmanspacher, T.~Filk, and H.~R\"omer.
\newblock Complementarity in bistable perception.
\newblock In H.~Atmanspacher and H.~Primas, editors, {\em Recasting Reality:
  Wolfgang Pauli's Philosophical Ideas and Contemporary Science}. Springer
  Verlag, 2008.

\bibitem{AertsGaboraSozzo}
D.~Aerts, L.~Gabora, and S.~Sozzo.
\newblock Concepts and their dynamics: A quantum-theoretic modelling of human
  thought.
\newblock {\em Topics in Cognitive Science}, 5:737--772, 2013.

\bibitem{AertsSozzoVeloz}
D.~Aerts, S.~Sozzo, and T.~Veloz.
\newblock Quantum structure in cognition and in the foundations of human
  reasoning.
\newblock {\em Arxiv 1412.8704}, 2015.

\bibitem{AertsDHooghe}
D.~Aerts and B.~D'Hooghe.
\newblock Classical logic versus quantum conceptual thought: Examples in
  economics, decision theory and concept theory.
\newblock {\em LNCS}, 5404:128--142, 2009.

\bibitem{PothosBusemeyer}
E.M. Pothos and J.D. Busemeyer.
\newblock A quantum probability model explanation for violations of rational
  decision theory.
\newblock {\em Proccedings of the Royal Society}, B276:2171--2178, 2009.

\bibitem{WangBusemeyer}
Z.~Wang and J.D. Busemeyer.
\newblock A quantum question order model supported by empirical tests of an a
  priori and precise prediction.
\newblock {\em Topics in Cognitive Science}, 5:689--1710, 2013.

\bibitem{QuantumConsciousness}
H.~Atmanspacher.
\newblock Quantum approaches to consciousness.
\newblock In E.~Zalta, editor, {\em Stanford Encyclopedia of Philosophy},
  updated 2011.

\bibitem{RoemerJacoby2011}
H.~R\"omer and G.E. Jacoby.
\newblock {S}ch\"opfer, {S}ch\"opfung, {S}ch\"opfertum, {T}alk {O}ffenburg
  2011.
\newblock {\em To appear, available on the author's homepage}.

\bibitem{Roemer2011}
H.~R{\"o}mer.
\newblock Verschr\"ankung (2008).
\newblock In M.~Knaup, T.~M\"uller, and P.~Sp\"at, editors, {\em
  Post-Physikalismus}, pages 87--121. Verlag Karl Alber, Freiburg i.Br., 2011.

\bibitem{Aertsetal}
D.~Aerts, T.~Durt, T.~Grib, B.~Van~Bogaert, and A.~Zapatrin.
\newblock Quantum structures in macroscopic reality.
\newblock {\em Internationanal Journal of Theoretical Physics}, 32:489--498,
  1993.

\bibitem{GA2006}
P.~beim Graben and H.~Atmanspacher.
\newblock Complementarity in classical dynamical systems.
\newblock {\em Foundations of Physics}, 36:291-- 306, 2006.

\bibitem{BFA2013}
P.~{b}eim Graben, Th. Filk, and H.~Atmanspacher.
\newblock Epistemic entanglement due to non-generating partitions of classical
  dynamical systems.
\newblock {\em International Journal of Theoretical Physics}, 52:723--734,
  2013.

\bibitem{Kim1993}
J.~Kim.
\newblock {\em Supervenience and Mind}.
\newblock Cambridge University Press, 1993.

\bibitem{RW2011}
H.~R{\"o}mer and H.~Walach.
\newblock Complementarity between phenomenal and physiological observables
  ({O}ct 2008).
\newblock In H.~Walach, S.~Schmidt, and W.B. Jonas, editors, {\em Neuroscience,
  Consciousness and Spirituality, ISBN 978-94-007-2078-7, DOI
  10.1007/978-94-007-2079-4}, pages 97--107. Springer publ. Comp., 2011.

\bibitem{Kim2003}
J.~Kim.
\newblock Blocking causal drainage and other maintainance chores with mental
  causation.
\newblock {\em Philosophical and Phenomenological Research}, 67:151--176, 2003.

\end{thebibliography}

\end{document}